\documentclass[%
aip,
 amsmath,amssymb,
 floatfix,
 reprint,%
author-year,%
floatfix,
]{revtex4-1}

\usepackage[T1]{fontenc}
\usepackage{graphicx}% Include figure files
\usepackage{dcolumn}% Align table columns on decimal point
\usepackage{bm}% bold math
\usepackage{todonotes}
\usepackage{relsize}
\usepackage{lmodern}

\graphicspath{{figs/}}

\usepackage{setspace}

    \newcommand{\sssection}[1]{%
      \subsubsection[#1]{\raggedright\normalsize\sffamily\itshape\bfseries #1}}
    \let\oldthesubsubsection\thesubsubsection
    \renewcommand{\thesubsubsection}{\begingroup \normalsize\sffamily\itshape\bfseries \oldthesubsubsection \endgroup}
\newcommand{\celerity}{\ensuremath{\upsilon}}
\newcommand{\ceff}{\ensuremath{c_\text{eff}}}
\newcommand{\dd}{\ensuremath{\text{d}}}

\begin{document}
\title[\textsf{Estimating winds from infrasound observations}\hfill \emph{E-print typeset by the authors}]{Estimating tropospheric and stratospheric winds using infrasound from explosions}% Force line breaks with \\

\author{Erik M\aa rten Blixt}
\email{marten.blixt@norsar.no}
\affiliation{NORSAR, Gunnar Randers vei 15, 2027 Kjeller, Norway}
\author{Sven Peter N\"{a}sholm}
\email{peter@norsar.no}
\affiliation{NORSAR, Gunnar Randers vei 15, 2027 Kjeller, Norway}
\author{Steven J.~Gibbons}
\affiliation{NORSAR, Gunnar Randers vei 15, 2027 Kjeller, Norway}
\author{L\"{a}slo G.~Evers}
\thanks{Also at: Royal Netherlands Meteorological Institute, R\& D Seismology and Acoustics, De Bilt, Netherlands}
\affiliation{Delft University of Technology, Applied Geophysics \& Petrophysics, Delft, Netherlands}
\author{Andrew J.~Charlton-Perez}
\affiliation{Department of Meteorology, University of Reading, Reading, UK}
\author{Yvan J.~Orsolini}
\affiliation{Norwegian Institute for Air Research, Norway}
\author{Tormod Kv\ae rna}
\affiliation{NORSAR, Gunnar Randers vei 15, 2027 Kjeller, Norway}
    
\date{27 June 2019}

\begin{abstract}
The receiver-to-source backazimuth of atmospheric infrasound signals is biased when cross-winds are present along the propagation path. %
Infrasound from 598 surface explosions from over 30 years in northern Finland is measured with high spatial resolution on an array 178 km almost due North. %
The array is situated in the classical shadow-zone distance from the explosions. %
However, strong infrasound is almost always observed, which is most plausibly due to partial reflections from stratospheric altitudes. %
The most probable propagation paths are subject to both tropospheric and stratospheric cross-winds, and our  wave-propagation modelling yields good correspondence between the observed backazimuth deviation and cross-winds from the ERA-Interim reanalysis product. %

We demonstrate that atmospheric cross-winds can be estimated directly from infrasound data using propagation time and backazimuth deviation observations. %
We find these cross-wind estimates to be in good agreement with the ERA-Interim reanalysis.
\end{abstract}

%Use showkeys class option if keyword %display desired
\maketitle

%\vspace{10pt}
%\begin{widetext}
\onecolumngrid
\begin{center}
\begin{spacing}{1.4}
\noindent\fbox{\begin{minipage}{.80\textwidth}{ \flushleft\textsf{\large This article has been accepted by The Journal of the Acoustical Society of America.
\newline
\noindent After it is published, it will be found at http://asa.scitation.org/journal/jas 
%\rule{0.5\textwidth}{0.5pt}
\newline
The current e-print was typeset by the authors and can differ in, e.g., pagination, reference numbering, and typographic detail.}}
\end{minipage}}
\end{spacing}
\end{center}
\par
\twocolumngrid

\section{\label{sec:intro}Introduction}
Infrasound are inaudible acoustic waves that can travel hundreds or thousands of kilometers in the atmosphere. %
The propagation of these waves is directly affected by wind and temperature conditions, which is particularly interesting because it provides a potential for 
remote sensing of the dynamics of the middle atmosphere, i.e.~the stratosphere and the mesosphere. %
Variations in the stratospheric conditions can propagate all the way down to the surface %
and an enhanced representation of the middle atmospheric dynamics in atmospheric model products is expected to enhance the skills of numerical weather prediction and climate forecasting systems at monthly timescales %
\citep{Baldwin2001, Baldwin2003, Charlton2007, karpechkoWMO2016, blanc2018toward, Pedatella2018}. 

The current study considers the use of infrasonic waves generated from man-made surface explosions to estimate the spatio-temporal average of a horizontal atmospheric wind component. %
Pioneering works on the exploitation of infrasound to probe the atmosphere include %
\citet{Groves1956}, \citet{Donn1972} and \citet{Rind1973}. %
More recent publications report on the development of inversion approaches to retrieve corrections to wind and temperature model profiles %
using using atmospheric infrasound recordings \citep{LePichon2005a,LePichon2005b, Drob2010inversion, Lalande2012infrasound, Assink2013estimation,Assink2019advances}. %
Moreover, several works report on monitoring the stratospheric polar vortex and the evaluation of sudden stratospheric warming forecasts using atmospheric infrasound datasets, see \citet{smets2019study,Smets2016} and the references therein. %

Infrasound arrivals are typically observed from tropospheric, stratospheric, and thermospheric waveguides. %
Mesospheric arrivals are more rare because of the negative temperature gradient at these altitudes. %
Knowing the celerity, \celerity{}, which is the ratio between horizontal great-circle source-receiver distance and travel time, can help when discriminating between arrivals from different atmospheric waveguides \citep{Nippress2014}. %

The atmospheric spatial and temporal structure of temperature and winds determine the availability of acoustic waveguides %
\citep{Georges1972, Garces1998, Drob2003}. %
It is the vertical temperature gradients and the winds along the horizontal line of propagation (the along-track winds or \emph{tail-winds}) that determine %
if a waveguide is present to return infrasound back down to ground level. %
The effective sound speed, \ceff{}, which is the adiabatic sound speed (proportional to the square-root of temperature) plus the horizontal tail-wind, is often used to quantify the presence of such waveguides. %

However, the wave propagation is also affected by the %
horizontal wind vector component normal to the line of propagation, the \emph{cross-wind}. %
This translates the medium of the propagating wavefront, %
making the source location appear as shifted sideways, as discussed thoroughly by \citet{Diamond1964} and \citet{Blom2017} among others. %
This is observed at the infrasound station as a deviation between true and measured backazimuth direction-of-arrival. 

In this paper we compare 598 observations of such deviations with cross-wind representations in the European Centre for Medium-Range Weather Forecasts (ECMWF) ERA-Interim atmospheric reanalysis product \citep{dee2011_era-interim} and assess how the cross-wind affects infrasound propagation. %
Based on insights gained, we then suggest and analyze a method to estimate atmospheric cross-winds directly from infrasound observations.

In Section \ref{sec:method}, we present the 30-year long time-series of 598 controlled and  well-characterized explosions in August and September, %
from which infrasound waves are propagating through the atmosphere.
Then the associated observations at a ground-based station at 178 km distance from the explosion site and the applied data processing are explained. %
This section also introduces the infrasound dataset, its processing, and the numerical wave propagation applied. %

We then discuss two approaches to assess cross-wind effects on infrasound propagation: one which assumes that the upper stratospheric cross-wind is dominating the backazimuth deviation effects, and another which takes the cross-wind component along the whole propagation path and altitude range into account.

In Section \ref{sec:results}, we demonstrate that a high correlation between the ERA-Interim reanalysis cross-wind and observed backazimuth deviation is found when the cross-wind along the whole path is taken into account. %

In Section \ref{sec:discussion}, we propose a straightforward approach to estimate the mean cross-wind directly from the infrasound observations. % 
We then compare this with numerical modelling using ray-tracing simulations through temperature and wind profiles retrieved from atmospheric reanalysis products. %
Section \ref{sec:summary} summarizes this work and presents an outlook on future research opportunities.

\section{\label{sec:method}Method}

\subsection{\label{ssec:groundtruthevents}Ground truth events}
A dense network of infrasound and seismic array stations is deployed in northern Fennoscandia and northwest Russia.  %
Operating for several decades, this network has recorded signatures from a large number of explosions resulting from mining activity and controlled explosions of obsolete military equipment \citep{Gibbons2010}. %
\citet{Gibbons2015} provides a thorough description of the array station network and describes event detection and characterization of seismo-acoustic events. %
Repetitive ground truth explosive events of anthropogenic origin have been registered in a database covering nearly 16000 explosions from 1987 up to present. %
For example, this database was exploited in \citet{Smets2015} to analyze one year of events from the Aitik open-pit mine in northern Sweden. %
In that paper, infrasound recorded at the IS37 infrasound station in northern Norway was interpreted in the context of wave propagation simulations through wind and temperature fields extracted from the ECMWF ensemble analysis product \citep{molteni1996ecmwf}. % 

Of the 14 characterized sites with repeating events, the military blast site of Hukkakero (67.94$^\circ$ N, 25.84$^\circ$ E), Finland, is of particular interest. %
Here, explosions with a yield of around 20 tons TNT equivalents are conducted yearly in August and September, repeating typically once a day with a similar yield every time. %
These strong explosions take place at the surface, with good coupling to both the ground and the atmosphere, producing clear and distinct seismic and infrasonic signals \citep{Gibbons2007}. %
This event-to-event repeatability and the long time coverage of events provide an opportunity to characterize the atmosphere by means of infrasonic probing. %
The current study focuses on 598 Hukkakero events with infrasound recorded on a single station.

\subsection{\label{ssec:obs}Observations}
\begin{figure}[tbhp]
    \centering
    \includegraphics[width=\columnwidth]{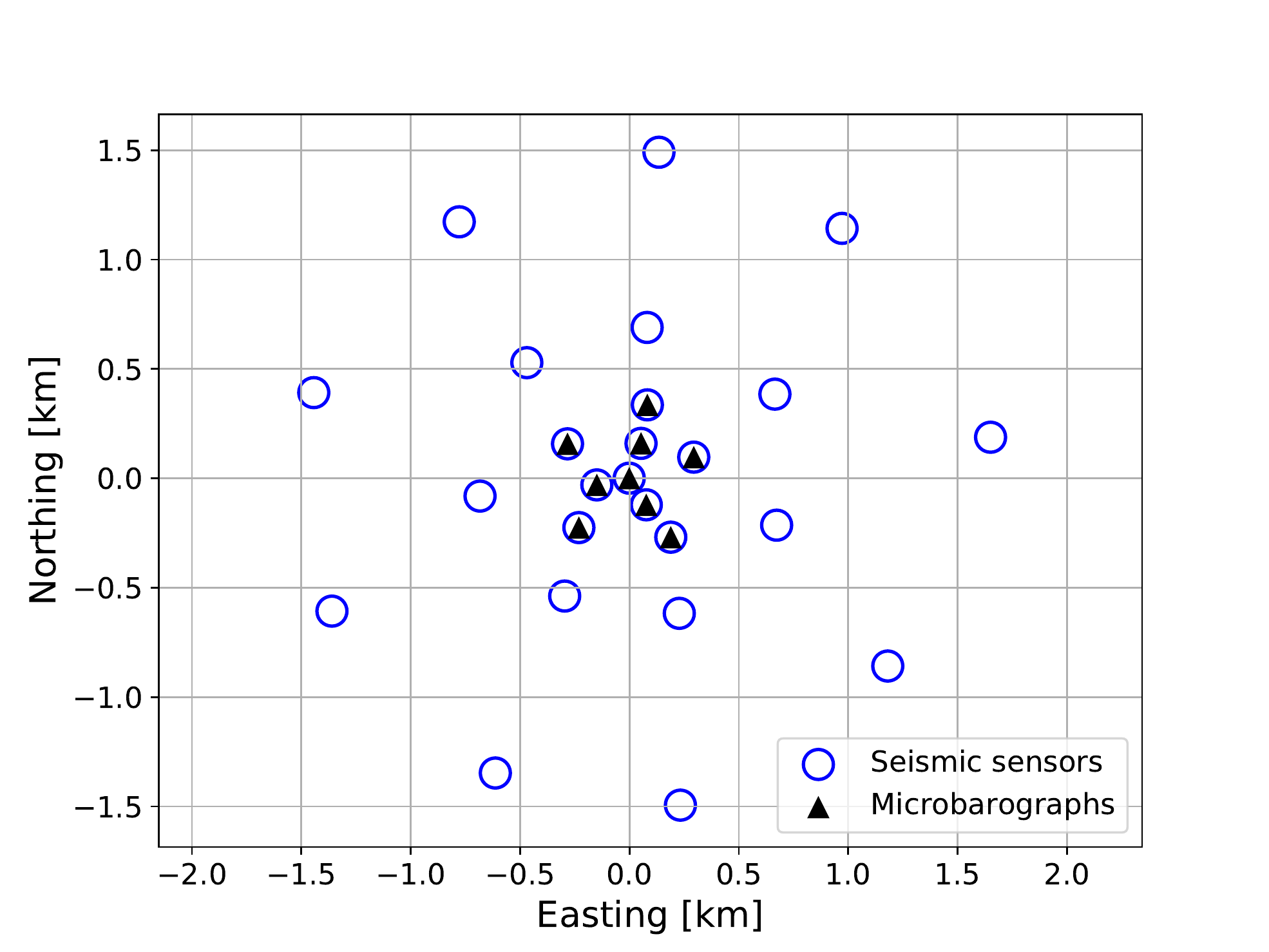}
    \caption{(Color online) 
    Array geometry of the ARCES seismic array (blue circles), with the microbarographs of the co-located infrasound array, ARCI, shown as black triangles.}
    \label{fig:geometry}
\end{figure}

The Hukkakero events are detected on several seismic and infrasound sensors in the European Arctic and beyond. %
In the current study, we report on data recorded at the ARCES/ARCI array  (Fig.~\ref{fig:geometry}) with co-located seismic and microbarograph 
sensors (69.53$^\circ$ N, 25.51$^\circ$ E) in Karasjok, northern Norway. The seismic array has the advantage of: %
1) Continuous observations for over 30 years; 2) Favourable large-array design with high azimuthal resolution. %
The nearly south-north alignment of this source-receiver pair %
makes the cross-wind components almost parallel to the eastward (zonal) winds, 
which have the largest magnitude of the middle atmospheric winds \citep{Drob2003}.

Each explosion generates distinct $P$ and $S$ phase signatures which have travelled through the solid earth. Around 10 minutes after these arrivals, %
a coupled atmospheric acoustic-to-seismic arrival is seen for 99\% of the events. %
This is the signature of the atmospheric infrasound wave that has travelled with the acoustic speed-of-sound, an order of magnitude lower than the speed of the seismic waves travelling through the solid earth,
which is then converted to ground-motion at the seismic array.

ARCES was deployed in 1987 with 25 nodes distributed within a 3 km aperture \citep{Mykkeltveit1990}. Fig.~\ref{fig:geometry} depicts the array geometry of ARCES, where the shortest inter-sensor distance is 145 m and the full aperture is 3114 m between its outermost nodes. %
Applying the \citet{Szuberla2004} uncertainty estimate approach, %
the ARCES array geometry is associated with a backazimuth uncertainty at less than 0.1$^{\circ}$ for a 95\% confidence interval, under the assumption of 0.05\,s infrasound phase arrival time uncertainty. %
In 2008, this seismic array was complemented with a co-located experimental microbarograph array, ARCI \citep{Evers2011}.

\begin{figure*}
    \centering
    \includegraphics[width=\textwidth]{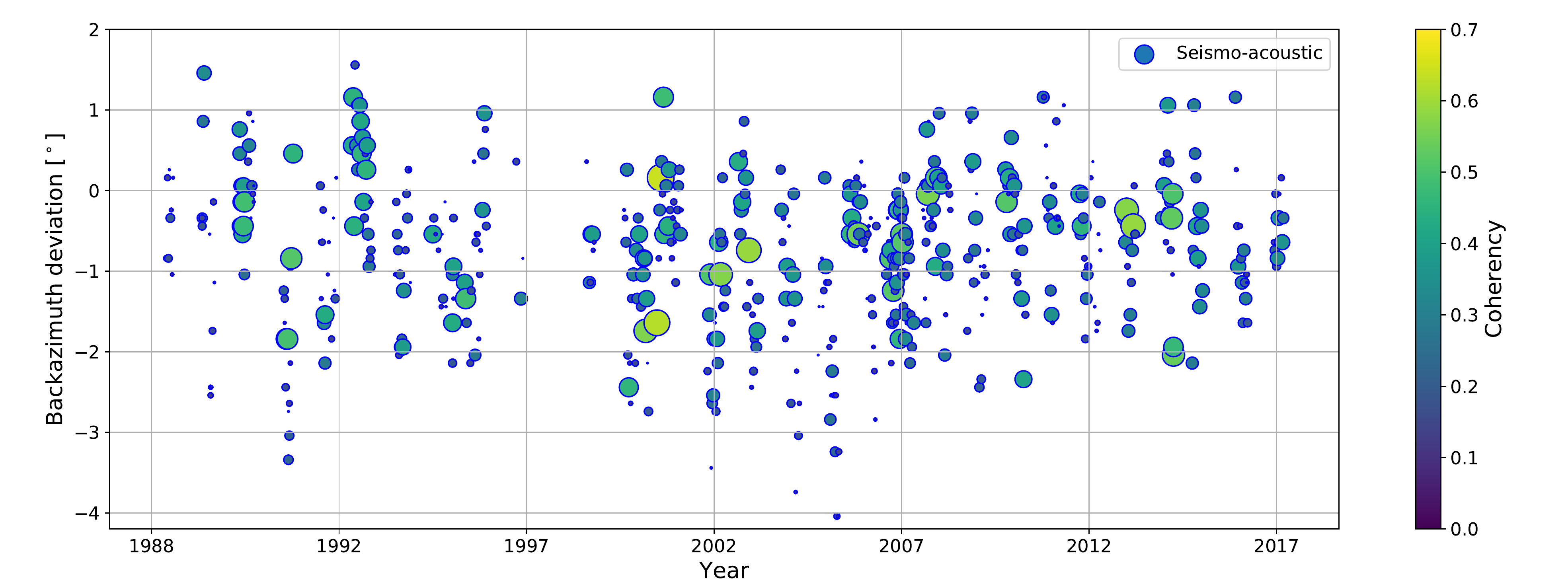}
    \caption{(Color online) Backazimuth deviation in infrasound arrivals from all Hukkakero explosions observed at the ARCES array from 1988 to 2017. %
    A negative backazimuth deviation means that the source appears to be shifted eastward from the true great circle direction from the station to the Hukkakero site. %
    Plot symbol size and colors are proportional to the average correlation between sensor traces associated with the event. Each event is represented by a single plot symbol which corresponds to the time instance with the highest average correlation between the sensor traces. % 
    The horizontal axis is discontinuous, because for each year only the days between August 10 and September 20 are plotted, as this is the date range within all explosions occur. %
    }
    \label{fig:timeazdev}
\end{figure*}
Fig.~\ref{fig:timeazdev} displays the full time-span of data observed at ARCES, covering more than 30 years. %
This figure displays the deviation in backazimuth from the true bearing towards Hukkakero (observed\,--\,true). %
 
Because of the long-term station operation, its stable and favourable large-aperture array geometry, %
and the strong signals from Hukkakero which generate appropriate pressure-to-ground-motion conversion, %
the current study only considers infrasound recorded at the ARCES seismic array. %

\subsection{\label{ssec:dataproc}Infrasound data processing}
\begin{figure}[tbhp]
    \centering
    \includegraphics[width=\columnwidth]{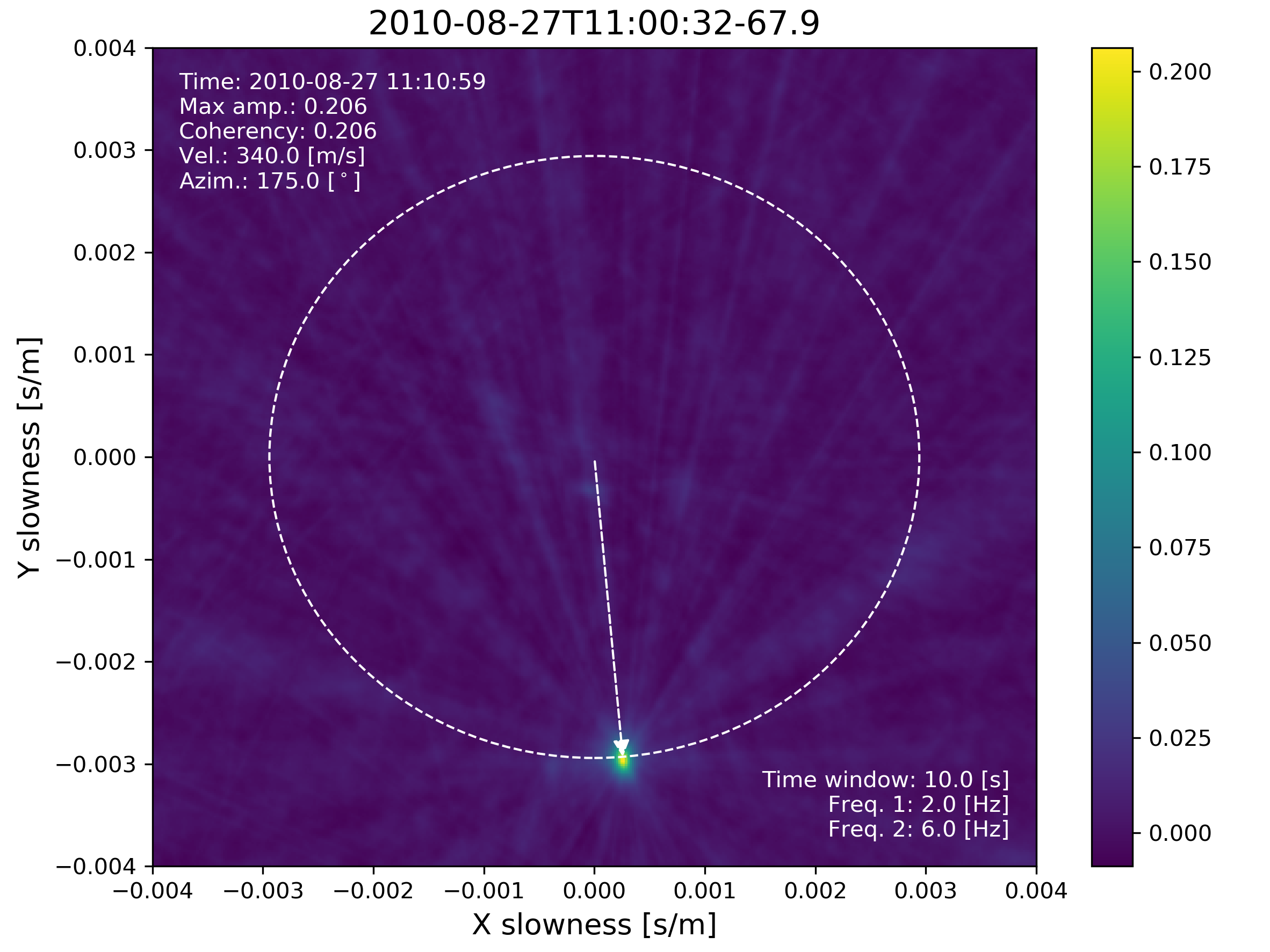}
    \caption{Slowness plot based on the cross-correlation traces for the Hukkakero explosion on 27 August 2010 as observed at the ARCES seismic array. The signal traces are filtered with a pass-band between 2 and 6 Hz. This event yields an average coherency of 0.206, and in the analysis that follows, events with a coherency below 0.2 have been ignored. Note that although this event barely passes the coherence criterion, the associated correlation peak is distinct with very weak sidelobes.}
    \label{fig:slowness}
\end{figure}

An $f$--$k$ analysis is performed on cross-correlations traces formed between all sensor trace pairs \citep{Brown2002, Gibbons2015}. %
The wavefront parameters (backazimuth and trace velocity) are determined from the slowness space coordinate where the average cross-correlation value between all sensor traces has its maximum. %
This analysis is done on 10--12\,s long-time 
windows, evaluated at 4\,s intervals. %
The seismo-acoustic data is filtered using a Butterworth pass-band filter to 2--6 Hz, in order to reduce the low-frequency noise microseismic and microbarometric noise contributions. 
The shortest inter-sensor distance is roughly double the wavelength of a 6 Hz infrasound wave. However, the array geometry of ARCES (Fig.~\ref{fig:geometry}) effectively suppresses sidelobes, and Hukkakero explosions typically yield a distinct slowness grid peak in the $f$--$k$ analysis, as exemplified in Fig.~\ref{fig:slowness}.

\begin{figure}[tbhp]
    \centering
    \includegraphics[width=\columnwidth]{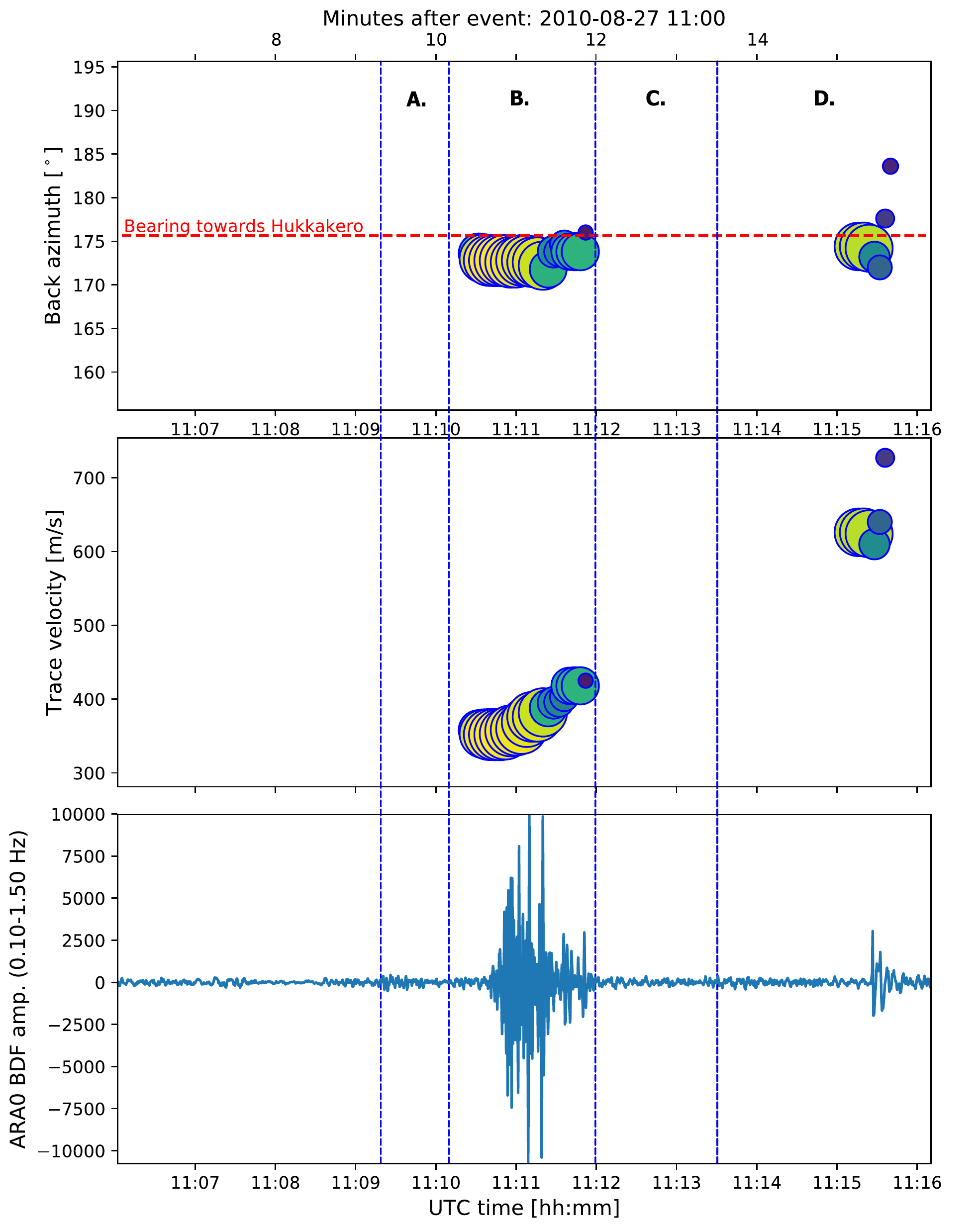}
    \caption{(Color online) Example of an infrasound signal from an explosion in 2010. The horizontal axis is UTC time and minutes after 
    explosion in all three panels. Lower panel is the pressure recorded at one of the microbarograph nodes in the ARCI array, the middle panel shows the trace, or apparent, velocity and the upper panel is the observed backazimuth of the infrasound signal in degrees East of North.
    The vertical dashed lines indicate the assumed tropospheric (A.), stratospheric (B.), mesospheric (C.) and thermospheric (D.) celerity ranges 
    ($310 - 341$\,m/s, $260 - 310$\,m/s $230 - 260$\,m/s and $< 230$\,m/s, respectively), based on the models of 
    \citet{Nippress2014,Modrak2010}, and \citet{Whitaker2008}.
    The size and color of the plot symbols are proportional to the coherency of the signal.}
    \label{fig:processing}
\end{figure}

Fig.~\ref{fig:processing} displays the trace and array parameters associated with one single Hukkakero event blast. %
Here, the different phase arrivals are clearly separated, and it is evident that there is significant  temporal variation in trace velocity and backazimuth within and between the arrivals. %
The upper panel shows the backazimuth of the detections, with the bearing towards Hukkakero drawn in a dashed red line for reference, displaying a backazimuth deviation of a couple of degrees from Hukkakero.
The middle panel displays the trace velocity.  %
The stratospheric signal increases its trace velocity during the later part of the stratospheric celerity range, indicating that it arrives at a steeper angle. %

We pick the backazimuth deviation (observed backazimuth minus great circle backazimuth) estimated at the time of highest coherency within the stratospheric celerity range to represent the backazimuth deviation and travel time of each event. %

The variability in backazimuth associated with the stratospheric arrival is typically within one degree, as exemplified in Fig.~\ref{fig:processing}. 
While the confidence in the backazimuth measurement is high for the large-aperture ARCES array using the \cite{Szuberla2004} approach -- only 0.1$^\circ$ using a 95\% confidence interval -- there is additional uncertainty due to other factors than time-delay uncertainty. %
Effects contributing to backazimuth measurement uncertainty include coherence loss, local meteorological and turbulence conditions, as well as range-dependent atmospheric variability. A thorough dissemination of these factors are beyond the scope of this article, but in the following we empirically quantify the total uncertainty: 
For each event, we assess the width of the main lobe in the slowness grid, as exemplified in  Fig.~\ref{fig:slowness}. %
A Gaussian function is the fitted to the cross-correlation values calculated over the slowness grid, along the constant-slowness circle (indicated with a white dashed circle in Fig.~\ref{fig:slowness}). Then we use the standard deviation of the fitted Gaussian as a backazimuth uncertainty estimate in degrees. %
This yields a value around $1.0 - 1.5 ^{\circ}$ for the events, as shown using vertical error bars in  Figs.~\ref{fig:wc-azdev-point} and \ref{fig:tt_wc-azdev-point}.

\subsection{\label{ssec:propagation}Infrasound propagation}
We use the ART2D engine \cite{hedlinwalker2013,walker_art2d}, to trace acoustic rays within temperature and wind models taken from ERA-Interim atmospheric reanalysis products. %
The ray-theoretical approximation is in general considered valid for a smooth medium with sound speed and wind gradient length scales being larger than the infrasound wavelength, which is $\sim$300\,m at 1\,Hz. %
In the current study we consider 1D atmospheric models, which is normally a valid assumption at regional source-receiver distances \citep{Assink2012}. %
See \citet{Blom2017} for a thorough discussion on infrasound ray tracing. %

The great circle distance between Hukkakero and ARCES is 178 km, and we apply conventional ray-tracing simulations through ERA-Interim atmospheric reanalysis temperature and wind fields to assess the size of the classical shadow zone (see, e.g., \citet{Wegener1925, Whipple1935}). % is done out to 280 km ground distance from Hukkakero. %
Within 280 km distance, we find that only 17\% of the stratospheric infrasound rays are refracted back down to the ground, and there is only one event with rays hitting the ground within 200 km range (at 192 km). %
The modelling hence confirms that this station is within the shadow zone range from Hukkakero. %

Although this station is located in the shadow zone, 99\% of the explosions are clearly observed in the infrasound data. %
This supports the hypothesis that partially reflecting structures not resolved by the reanalysis, are present at stratospheric altitudes and that these can guide the propagating acoustic wave down to the station. %
Such layered structures can be attributed, e.g, to wind shear or internal gravity wave perturbations to the smoother background atmospheric profile \citep{Chunchuzov2019internal,Kulichkov2010, Green2018, chunchuzov2014modeling}.%
 
\begin{figure}[tbhp]
    \centering
    \includegraphics[width=\columnwidth]{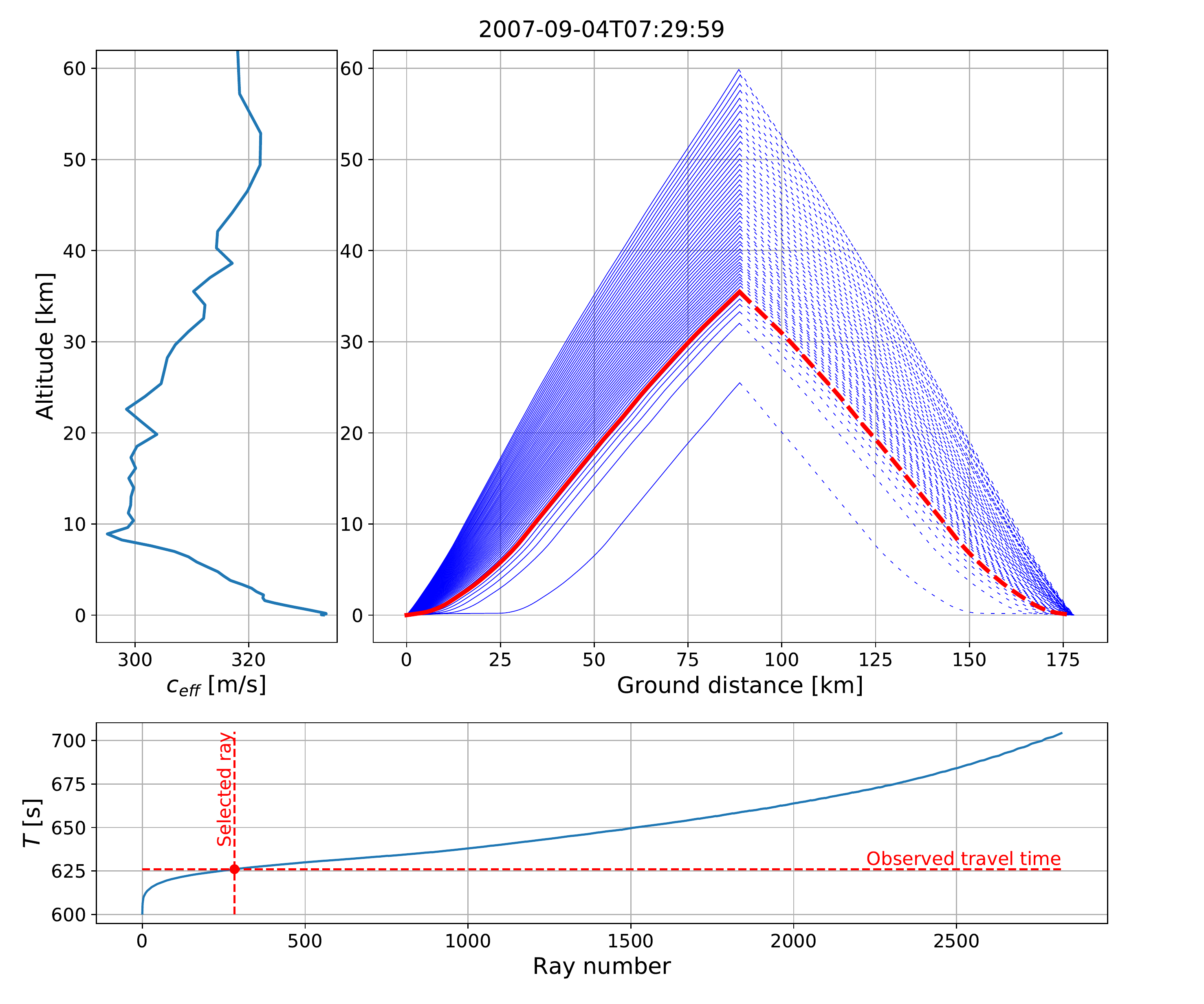}
    \caption{(Color online) Partially reflected ray paths (upper right panel), with different launch angles, for explosion on 4 September 2007. %
    The dashed lines correspond to mirrored ray paths after partial reflection at the midpoint between explosion and the array at $178$ km range. The lower panel shows the modelled travel time for each ray. The ray with the smallest difference between modelled and measured travel time (626\,s in this example) is selected to represent this event (thick red line). %
    The upper left panel shows the effective sound speed as a function of altitude for this 1D atmospheric model.}
    \label{fig:partial_reflection}
\end{figure}

Fig.~\ref{fig:partial_reflection} displays our numerical modelling of ray paths involving a stratospheric partial reflection or scattering at the great-circle midpoint between the explosion site and the station for an explosion on September 4, 2017. %
For each event, a 1D atmospheric model is extracted from the ERA-Interim reanalysis product. %
The temperature and wind profiles are read at the ERA-Interim gridpoint closest to the great circle midpoint between source and receiver, and closest in time to the event origin time plus the approximate travel time to midpoint (5 minutes). %
Our ray-tracing simulations is set up with a dense fan of infrasound rays launched from Hukkakero with inclination angles ranging from $0^\circ$ to $85^\circ$. %
We trace these rays to the midpoint between site and station, and then model stratospheric partial reflections by mirroring all simulated upward propagating rays to fall back to the surface. This is similar to the approach displayed, e.g., in \citet{chunchuzov2015characteristics} (Fig.\,6) and in \citet{Green2018} (Fig.\,18). %
This way, we generate a set of eigenrays connecting the event site with the station. %
Each of these eigenrays is associated with a traveltime, and we select the ray with travel time closest to the observed travel time. 
The top-right panel of Fig.~\ref{fig:partial_reflection} highlights the selected ray with a thick red line for the analyzed event. %
There is a good agreement between observed and modelled travel time, typically with a discrepancy of less than 0.1\,s. 
Both modelling and data hence support the assumption that partially reflecting or scattering stratospheric structures can explain the observed infrasound arrivals recorded in the shadow zone.

\subsection{\label{ssec:crosswind}Cross-wind estimation}
We define the cross-wind $W_c$ to point along the $X$ axis in a 
right-handed coordinate system that has its $Y$-axis aligned with the great circle connecting the explosion site and the station, positive in the along-track direction, with $Z$ positive upwards.  %
So, for a wave moving northward from Hukkakero towards ARCES, a positive cross-wind is directed eastward. %

\sssection{\label{sssec:local_strat}Assessing whether stratospheric cross-wind effects dominate}
For a typical diffracted ray situation outside of the shadow-zone, the most coherent waves are often assumed to return from the stratospheric altitude where the 
ratio between the effective sound speed on ground and at altitude reaches a value at around one \cite{Evers2010}. %

Assuming that the infrasound waves spend the most significant part the propagation time within the return height altitude range, %
implies that the cross-winds at these altitudes would have the greatest effect on infrasound propagation. %
We test this hypothesis for the current dataset by looking at a \emph{single-point} stratospheric cross-wind estimation at the return height, which is derived from ERA-Interim reanalysis, evaluated at the midpoint between source and receiver, for each event. The resulting cross-wind estimate is plotted versus the observed backazimuth deviation, in Fig.~\ref{fig:wc-azdev-point}

\sssection{\label{ssec:strat_and_trop}Combined stratospheric and tropospheric cross-wind effects}

In contrast to the estimate of the stratospheric wind contribution above, the cross-wind effect along the propagating infrasound can be estimated
by the travel time weighted mean cross wind along the propagation path, defined as
\begin{equation}
    \label{eq:mean_wc}
       W_{c,T}\equiv \dfrac{1}{T}\int_{0}^{T} W_c(t)\, \dd t,
\end{equation}
where $T$ is the total travel time.
For a given ray path, this mean cross-wind can be extracted from the reanalysis as a weighted sum of cross-winds, similar to the approach in \citet{Diamond1964}, and is plotted in Fig.~\ref{fig:tt_wc-azdev-point}. 
\section{\label{sec:results}Results}
\begin{figure}[tbhp]
    \centering
    \includegraphics[width=\columnwidth]{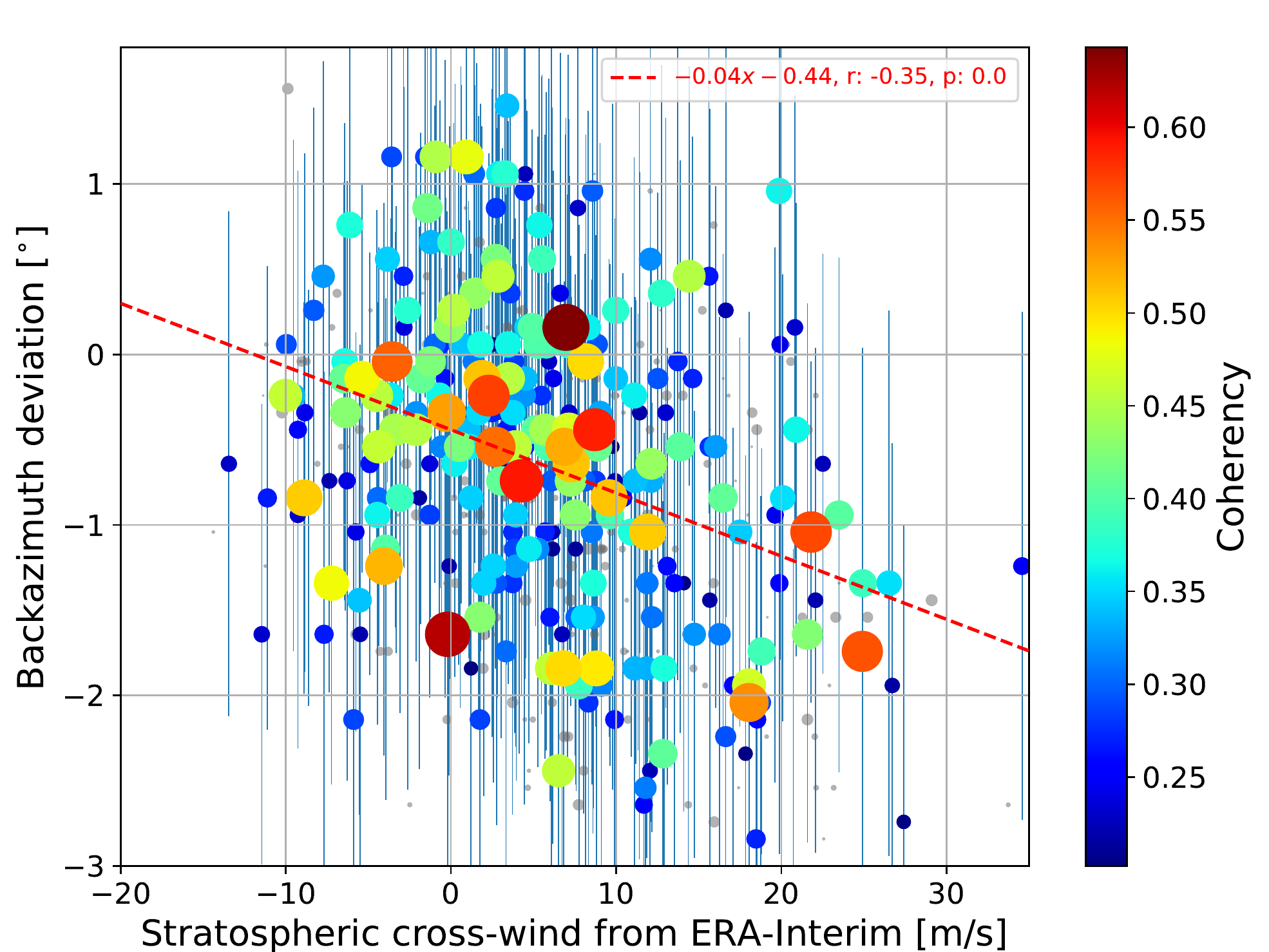}
    \caption{(Color online) Cross plot between the \textit{single-point} stratospheric cross-wind, at assumed return height, and the observed backazimuth deviation. Plot symbol size and colors are proportional to the observed infrasound correlation. %
    The red dashed line displays the linear regression data fit.}
    \label{fig:wc-azdev-point}
\end{figure}

\begin{figure}[tbhp]
    \centering
    \includegraphics[width=\columnwidth]{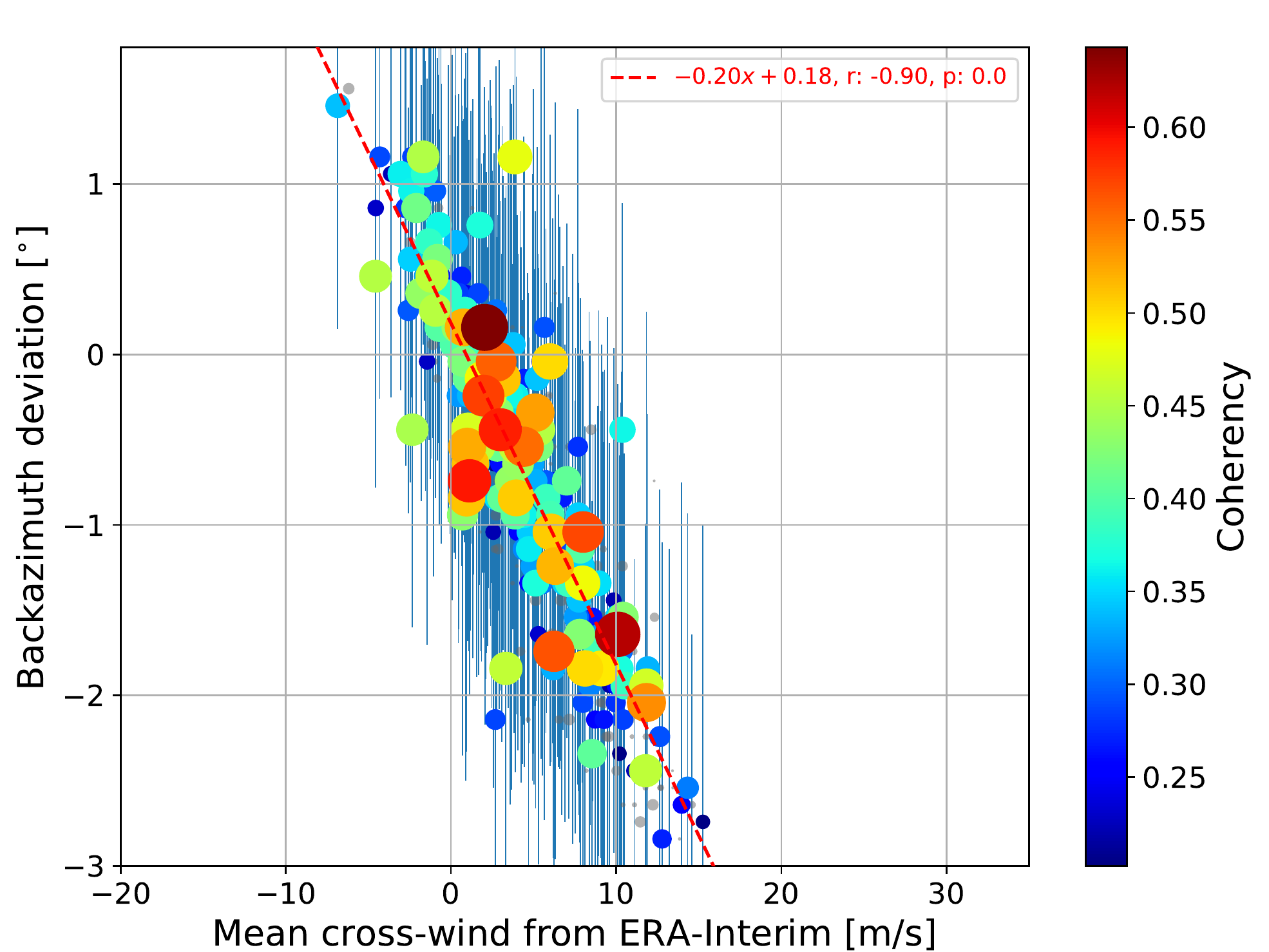}
    \caption{(Color online) Cross plot of the  travel time weighted mean cross-wind along a partially reflected ray, versus the observed seismo-acoustic backazimuth deviation. Size and color of the plot symbols are proportional to
    the observed infrasound coherency.}
    \label{fig:tt_wc-azdev-point}
\end{figure}

First, we apply the method described in Section \textmd{\ref{sssec:local_strat}} to assess whether stratospheric cross-winds are dominating the effects on backazimuth deviation. The horizontal axis of Fig.~\ref{fig:wc-azdev-point} displays the stratospheric cross-wind 
plotted against the backazimuth deviation in the infrasound data. %
In Fig.~\ref{fig:tt_wc-azdev-point}, the travel-time weighted mean cross-wind, as described in Section \ref{ssec:strat_and_trop}, is plotted against the measured backazimuth deviation. %

In the linear regression fit, plotted as a dashed red line in both Fig.~\ref{fig:wc-azdev-point} and \ref{fig:tt_wc-azdev-point}, we disregard detections with coherency less than 0.2, and points with lower 
coherency than this are colored gray. %
The linear function and the associated fit diagnostics are shown in red text in each figure. %
The scattered cloud of data points in Fig.~\ref{fig:wc-azdev-point} has a low correlation coefficient, $r$, and a $p$-value, $p$, of zero. 
A $p$-value of zero indicates that the linear data trend is likely.

Fig.~\ref{fig:tt_wc-azdev-point} shows less scatter, with the data points aligned along the linear regression fit with a slope of $-0.20\,\deg\,$s/m. 
The regression fit correlation value, $r$, is significantly increased from $-0.35$ to $-0.90$. Also, a visual comparison between these two figures confirms the augmented linear relation between infrasonic backazimuth and ERA-Interim cross-winds.

\section{\label{sec:discussion}Discussion}
The physical mechanism explaining the negative correlation in Fig.~\ref{fig:wc-azdev-point} and, predominantly, Fig.~\ref{fig:tt_wc-azdev-point} is that a positive cross wind (almost due East for this source\,--\,receiver configuration) will seemingly shift the source eastwards as seen from the seismic array, which is situated North of the explosion site. %
This creates an anti-clockwise (negative) deviation in backazimuth. %
Following \citet{Diamond1964}, this backazimuth deviation can be calculated from the horizontal great circle distance, $\celerity \cdot T$, that an infrasound wave has travelled during the time $T$,  %
where \celerity{} denotes the celerity. %
During this propagation time, the cross-wind translates the medium and the propagating wave laterally a distance $\int^{T}_0\: W_{c}(t)\, \text{d}t$.  %
From this follows that the observed backazimuth deviation $\Delta \theta$ is related to the cross-wind $W_c$ as
\begin{equation}
    \label{eq:diamond}
    \tan \Delta \theta = - \dfrac{1}{\celerity\, T}\int_{0}^{T} W_c(t)\, \dd t =  -\frac{ W_{c,T} }{\celerity},
\end{equation}
where $W_{c, T}$ is the travel-time weighted mean cross-wind, defined in Eq.~\ref{eq:mean_wc}.  %

For backazimuth deviations less than around 10$^{\circ}$, Eq.~\ref{eq:diamond} simplifies to $\Delta \theta \approx -\frac{1}{\celerity}W_{c,T}$, %
so in a cross plot of cross-wind versus backazimuth deviation, we expect the events to be aligned along a straight line % 
with a slope inversely proportional to average celerity. %
The mean travel time of all recorded events is $637.5$\,s, giving a mean celerity of $279$\,m/s. %
Following Eq.~\ref{eq:diamond}, this corresponds to a slope of $-0.21 \deg\,$s/m, which is consistent with the linear regression slope in Fig.~\ref{fig:tt_wc-azdev-point}. %

Fig.~\ref{fig:wc-azdev-point} exhibits significant scatter in the data. %
The linear regression features only a minor negative slope ($-0.04$ $\deg\,$s/m), which is far from what is expected from the mean travel time.
So applying the {\it single-point} method on this dataset, yields no convincing relation between reanalysis cross-winds and observed backazimuth deviation. %
Possible explanations of this low correlation include: 
    1) The stratospheric cross-wind, estimated from the ERA-Interim atmospheric reanalysis, is incorrect;
    2) The stratospheric cross-wind alone is not dominating the backazimuth deviation effect.

Since there is good agreement between measured, and modelled, travel time and backazimuth deviation,
explanation 1) is not likely. %
This observation also supports explanation 2), because the return height found in the traced infrasound rays is often lower than 
the altitude where $\ceff$ reaches its ground level value.%

We underline that, although the cross-wind estimation analysis in the current work is made on a large dataset of events, the stratospheric zonal winds between Hukkakero and the station are weak in August and September: typically below 10\,m/s. %
During this period of the year, the stratospheric wind climate typically reverses from its westward
summer pattern, and instead the eastward stratospheric polar vortex winter pattern develops. %
In comparison, the tropospheric cross-wind contributions are hence significant for these explosions. %
By contrast, the stratospheric winds in winter are generally much stronger: the zonal average in January at 70$^\circ$N is more than 40\,m/s \citep{waugh2017whatispolarvortex}, and it is not uncommon that January stratospheric winds speeds exceed 100\,m/s at high latitudes. %

The weak stratospheric winds, and the ray paths returning from below the assumed refractive return height support explanation 2) above.

In contrast to Fig.~\ref{fig:wc-azdev-point}, the slope at $-0.20 \deg$\,s/m in Fig.~\ref{fig:tt_wc-azdev-point} is very close to the estimate from 
\eqref{eq:diamond}. %
This confirms that the observed backazimuth deviations can be well explained by the mean cross-wind effect along a partially reflected acoustic ray, traced through an ERA-Interim atmospheric reanalysis. %

\begin{figure}[tbhp]
    \centering
    \includegraphics[width=\columnwidth]{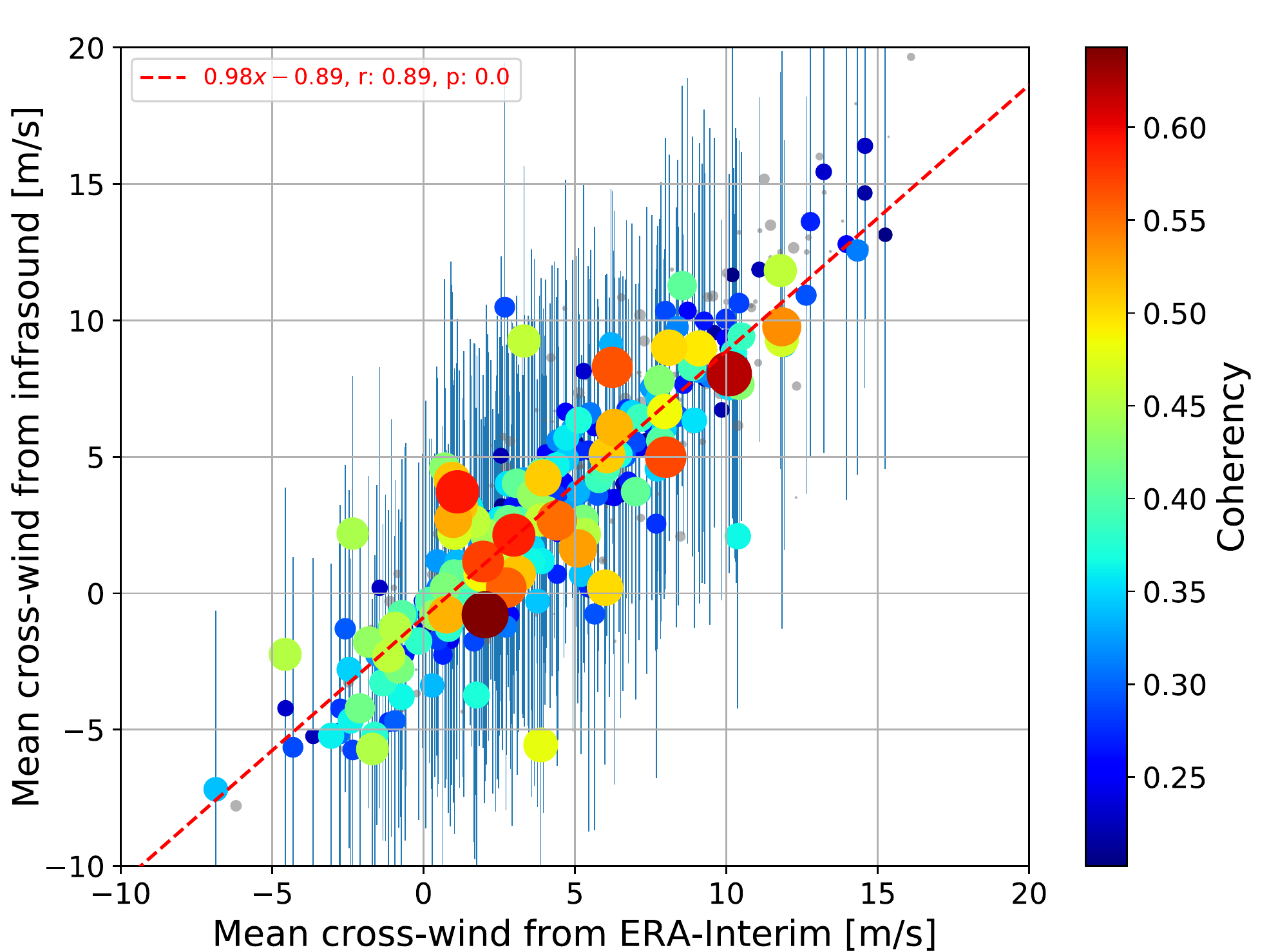}
    \caption{(Color online) The average cross-wind estimated from ERA-Interim atmospheric reanalysis (horizontal axis) against average cross-wind from 
    infrasound observations (vertical axis) for the 598 explosions. %
    Size and color of the plot symbols are proportional to the observed coherency between the sensor traces.}
    \label{fig:infrawinds}
\end{figure}

Noting that both the backazimuth deviation $\Delta \theta$, and the celerity $\celerity$, can be measured in the infrasound data, given that we know the origin time and position of the explosion
from seismic data that has propagated through the solid earth, %
we find an opportunity to measure the travel-time weighted average cross-wind $W_{c,T}$ for each event. This is done using relation \eqref{eq:diamond}, which after straightforward reorganization gives the average cross-wind as
\begin{align}
W_{c,T} = -\celerity \tan \Delta\theta. \label{eq:wct}%
\end{align}
This cross-wind estimate is plotted in Fig.~\ref{fig:infrawinds} for each event. %
Here we note that there is a clear correlation between the average cross-wind measured from the  infrasound data and the average cross-wind extracted from the ERA-Interim atmospheric reanalysis: the statistical analysis results in a  %
linear regression slope close to one, with an associated correlation coefficient of $0.89$. %

The cross-wind estimate uncertainty displayed in the error bars of Fig.~\ref{fig:infrawinds} is calculated using error propagation based on a linearization of Eq.\ \eqref{eq:wct} and the uncertainties in backazimuth deviation and travel-time: %
The contribution related to travel-time becomes proportional to $\partial T / T$, while the contribution related to   
backazimuth becomes proportional to $\partial \left(\Delta \theta\right)$, where $\partial$ denotes the uncertainty of each variable. % 
Assuming the travel-time uncertainty to be high, e.g., $\partial T=10$ s, we find that the contribution from backazimuth deviation still dominates and that the travel-time uncertainty contribution to the average cross-wind measurement is negligible. %
%Because of this, the error bars in Fig.~\ref{fig:infrawinds} are of indistinguishable length for $\partial T$ = 1 s and $\partial T$ = 10 s. However, a timing error of 1 s has been used in Fig.~\ref{fig:infrawinds}. 

Individual event deviation from the linear fit shown in Fig.~\ref{fig:infrawinds} %
can naturally be due to observation uncertainty, as discussed in the final paragraph of Section \ref{ssec:dataproc}. %
%, which can be larger than the theoretical uncertainties \citep{Szuberla2004}, because of coherence loss, local meteorological conditions at the receiving array, and atmospheric variability. 
We also note that the effective sound speed can differ with up to 30 m/s from ECMWF reanalysis due to gravity wave effects and momentum flux \citep{assink2014evaluation} and that this unresolved atmospheric variability could also affect the cross-wind. However, the small scale nature of such mechanisms suggests that the associated effects would be less prominent in average cross-winds over the regional source-receiver distance in the current study. %
Moreover, when the average stratospheric and tropospheric cross-winds are weak, as is typical especially for stratospheric altitudes in this August and September dataset, the observational uncertainty becomes more significant in relation to the cross-wind average. %
On the other hand, we would expect the potential in contributing to atmospheric model wind representation enhancement to be greater for events where there is difference between the re-analysis and the infrasound-based estimate of the cross-wind. %

We note that recorded traces from partially reflected infrasound can also be analyzed to probe the vertical profiles of layered wind velocity inhomogeneities \citep{chunchuzov2015reproducing, chunchuzov2015characteristics,chunchuzov2015study}, and that such approaches may also be applied to the current dataset.

Moreover, the presented results implicitly confirm that infrasound-based location estimates of atmospheric explosions are enhanced if cross-winds extracted from atmospheric model products are used to correct for backazimuth deviations as suggested by \citet{Garces1998} and successfully applied by \citet{Evers2007} as well as \citet{Evers2007infrasonic}. %

\section{\label{sec:summary}Summary and outlook}
We have analyzed 30 years of infrasound observations of ground truth explosions in northern Fennoscandia. %
Despite that the recording station, ARCES, is located in the classical shadow zone range for infrasound emitted at Hukkakero, 99\% of the explosions are detected. The observed infrasonic arrivals can be explained by partial reflections from fine-scale inhomogeneities in the stratosphere midway between source and station. %

The infrasound backazimuth deviation in the arrivals is shown to be well explained by cross-wind effects on the propagating wavefront. %
However, this deviation cannot be explained by the stratospheric cross-winds only and we need to consider the %
averaged cross-winds along the full propagation path and altitude range. %

We analyze a straightforward approach to estimate a snapshot of the spatially averaged cross-wind along the ray path from the combination of two infrasound arrival parameters: %
the observed travel time and the backazimuth deviation, see  Eq.\,\eqref{eq:wct}. %
Our wind estimates agree well with the ERA-interim reanalysis product, %
hence demonstrating a potential to infer atmospheric dynamics from infrasound observations alone. %

The applied data analysis methodology is not limited to stratospheric arrivals, but it can also implemented  
on tropospheric and thermospheric arrivals,
allowing for sampling of different layers of the atmosphere.

The method should also be extended to allow for the use of data from continuous sources such as microbaroms and volcanoes. %
A continuous sampling of the atmosphere is of great value, but this would require both a reliable estimate of the source location and a well-founded travel-time measurement which is less straightforward for continuous sources. %

Future work related to our study could also include assessments of the robustness of the average cross-wind inversion, e.g., by performing comprehensive full-waveform simulations where both measurement errors and realistic atmospheric model variations are taken into account.

Moreover, we expect upcoming studies to consider also data-based estimates of the average effective sound speed along the propagation path, and to assess this in the context of the atmospheric model products in a similar manner as the crosswind is studied in the current work.

A long-term  objective is to probe large-scale middle atmospheric winds by combining cross-wind components estimated at stations in a network which observe the same explosions. %
This work presents a first step towards achieving this objective.  %

We suggest that infrasound measurements can be used as independent observations to constrain atmospheric model products, and that future works will provide frameworks for infrasound data assimilation. %

\section*{Acknowledgments} 
The research leading to these results was partly supported by the project ``Middle Atmosphere Dynamics: Exploiting Infrasound Using a Multidisciplinary Approach at High Latitudes'' (MADEIRA), funded by the Research Council of Norway basic research programme FRIPRO/FRINATEK under contract number 274377. %
We also acknowledge support from the European Union under the Horizon 2020 infrastructure design study project ARISE2 under Grant Agreement Number 653980 \citep{blanc2019middle,blanc2018toward}, as well as NORSAR institute internal funding. %
We are grateful to Kristoffer T. Walker for making the ART2D ray-tracing software openly available. %
We thank the two anonymous reviewers whose suggestions helped improve and clarify this manuscript.

\listoffigures
\end{document}